\documentclass[lettersize,journal]{IEEEtran}
\usepackage{amsmath,amsfonts}
\usepackage{algorithmic}
\usepackage{array}
\usepackage[caption=false,font=normalsize,labelfont=sf,textfont=sf]{subfig}
\usepackage{textcomp}
\usepackage{stfloats}
\usepackage{url}
\usepackage{verbatim}
\usepackage{makecell}
\usepackage{graphicx}
\usepackage{multirow}
\usepackage{float}     
\usepackage{siunitx}
\usepackage{graphicx}
\usepackage{booktabs} 
\usepackage{amsmath} 
\def\BibTeX{{\rm B\kern-.05em{\sc i\kern-.025em b}\kern-.08em
    T\kern-.1667em\lower.7ex\hbox{E}\kern-.125emX}}
\usepackage{balance}
\begin{document}
\title{A Grouped Sorting Queue Supporting Dynamic Updates for Timer Management in High-Speed Network Interface Cards}
\author{Zekun Wang, Binghao Yue, Weitao Pan, Jianyi Shi, Yue Hao
\thanks{
Manuscript created October, 2025; This work is supported by the National Key R\&D Program of China (2023YFB4405100, 2023YFB4405102).(Corresponding author: Weitao Pan.) 

Zekun Wang, Jianyi Shi, Yue Hao are with the School of Microelectronics, Xidian
University, Xi'an 710126, China (e-mail: zekunwang@stu.xidian.edu.cn; jyshi@mail.xidian.edu.cn; yhao@xidian.edu.cn).
Weitao Pan, Binghao Yue are with the State Key Laboratory of Integrated Service Networks, Xidian University, Xi'an 710071, China (e-mail: wtpan@mail.xidian.edu.cn;  23011210684@stu.xidian.edu.cn).
}}

\markboth{Journal of \LaTeX\ Class Files,~Vol.~18, No.~9, September~2020}%
{How to Use the IEEEtran \LaTeX \ Templates}

\maketitle

\begin{abstract}
With the hardware offloading of network functions, network interface cards (NICs) undertake massive stateful, high-precision, and high-throughput tasks, where timers serve as a critical enabling component. However, existing timer management schemes suffer from heavy software load, low precision, lack of hardware update support, and overflow. This paper proposes two novel operations for priority queues—update and group sorting—to enable hardware timer management. To the best of our knowledge, this work presents the first hardware priority queue to support an update operation through the composition and propagation of basic operations to modify the priorities of elements within the queue. The group sorting mechanism ensures correct timing behavior post-overflow by establishing a group boundary priority to alter the sorting process and element insertion positions. Implemented with a hybrid architecture of a one-dimension (1D) systolic array and shift registers, our design is validated through packet-level simulations for flow table timeout management. Results demonstrate that a 4K-depth, 16-bit timer queue achieves over 500 MHz (175 Mpps, 12 ns precision) in a 28nm process and over 300 MHz (116 Mpps) on an FPGA. Critically, it reduces LUTs and FFs usage by 31\% and 25\%, respectively, compared to existing designs.

\end{abstract}

\begin{IEEEkeywords}
Priority Queue, Timer Queue, Systolic Arrays.
\end{IEEEkeywords}

\section{Introduction}
\IEEEPARstart{I}{n} modern high-performance network architectures, timers are fundamental mechanisms widely used in protocol stack management, resource recycling, and reliability assurance.\\
They are indispensable for flow table management in Software Defined Networking (SDN), Queue Pair (QP) maintenance in Remote Direct Memory Access (RDMA), and retransmission in TCP Offload Engines (TOE), requiring both sufficient precision for protocol correctness and dynamic updates to adapt to network changes.\\
In SDN, the control plane and data plane are decoupled \cite{b1}. The control plane deploys forwarding policies through southbound interfaces, and these policies are implemented on switches in the form of flow rules [2]. OpenFlow, a standard southbound protocol, defines the expiration time of flow rules using two parameters: idle\_timeout and hard\_timeout. It also provides three flow table management mechanisms [3], namely: flow expiration timeouts, proactive flow entry deletion, and flow entry eviction. Common software switches (e.g., Open vSwitch [4]) mainly rely on CPUs and software timers to manage entry timeouts. To configure flow entry timeouts reasonably and achieve much higher flow table usage, [5] and [6] both propose methods combining dynamic timeout adjustment and active eviction strategies for the idle\_timeout type. The Timeout is updated whenever packets of the same flow arrive; when the flow table utilization rate exceeds a threshold, specific entries are deleted. Despite improved utilization, the software-based approach suffers from low timing accuracy and require software involvement in timeout updates, leading to high CPU and memory usage—with the CPU usage rate exceeding 80\% [6]. Thus, the overhead of software-based dynamic timer management is relatively high. \\
In RDMA technology, the Reliable Connection (RC) service is one of its fundamental service types, and timers constitute a critical component of this service [7]. The reliable transmission mechanism is managed at the granularity of QPs; a dedicated timer is configured for each QP to enable the retransmission mechanism.
As the volume of QP activities increases, RDMA performance degrades with the growth of connection count [8]. This limitation can be addressed by offloading most retransmission mechanisms to hardware NICs, which enhances connection scalability [9]. However, conventional hardware timer management typically adopts a one-cycle-check approach [10], [11]. A key drawback of this approach is that timing precision decreases as the scale of timers expands—making it difficult to apply in high-speed network scenarios.\\
One of the key guarantees for TCP reliable services is also the retransmission timer [12], which is used to detect response loss during data transmission. [13] indicates that reasonably setting the Retransmission Timeout (RTO) based on Round-Trip Time (RTT) helps address the link idleness caused by retransmissions of a large number of concurrent flows.
With the development of TOE technology that offloads TCP functions to NICs [14], collecting RTT information and configuring hardware retransmission timers at the hardware level will effectively improve retransmission efficiency. Moreover, finer timing granularity is also beneficial for resuming retransmission and accurately measuring time intervals. However, common SmartNICs usually lack built-in timers, and timers in TOE technology are often implemented at the software level in conjunction with smart NICs [15],[16].\\
From the typical application scenarios of timers discussed above, the following shortcomings in current large-scale timer management can be summarized:
\begin{itemize}
	\item{The overhead of software-based dynamic timer management is relatively high;}
	\item{The accuracy of hardware timers is limited by scale expansion;}
	\item{SmartNICs lack efficient hardware timers.}
\end{itemize}

Among the numerous timer management methods, priority queues feature scalability and timing precision independent of scale. However, this method relies on an incrementing external reference timer to set expiration times for target tasks. Constrained by the fixed bit-width of hardware, timer overflow is inevitable—which causes timing errors—and no optimal solution has been identified to date.
To implement a hardware timer queue supporting dynamic updates and resolve timing errors induced by overflow, we propose update operations and group sorting. The design is realized based on a hybrid architecture of a 1D systolic array and shift registers, and ultimately applied to flow entry timeout management. The GitHub link for the code and test environment will be provided in the final manuscript.\\
The main contributions of this paper are as follows:
\begin{itemize}
	\item{From a shift-based perspective, we analyze the implementation of basic operations. By combining and propagating these basic operations, we achieve an update operation that supports dynamic updates to existing elements in the queue without disrupting the sorting result;}
	\item{Propose a group sorting mechanism that supports dynamic modification of search logic during the enqueue process. By utilizing the timer’s Most Significant Bit (MSB) as a grouping flag, this approach ensures timing accuracy after overflow with low hardware overhead;}
	\item{A hybrid architecture of a 1D systolic array and shift registers is adopted to implement the hardware priority queue. The timing precision is configurable, reaching up to 12 \(ns\);}
	\item{Based on a 28nm process library, when supporting a 4K depth and 64K priorities, the design can operate at over 500 MHz with a throughput of up to 175 Mpps; when implemented on an FPGA, it can run at 345 MHz with a throughput of 116 Mpps.}
\end{itemize}

\section{Related Work}
For the management of large-scale timer queues, three common schemes exist: the one-cycle-check method, the ordered list method, and the multi-level queue method. Among these, the ordered list method further includes the timing wheel and priority queue. This section analyzes these three methods as well as the implementation of related hardware priority queues.

The one-cycle-check method traverse all timers within one timing precision interval, and their timing values are decremented by 1. When a timer drops to 0, the corresponding event is triggered. 
Since the traversal time is related to the scale of the timer, if a system contains a large number of timers, the timing precision will be severely compromised.
Jingzhao RDMA NIC [10] employs hardware timers for retransmission and adopts the traversal
\begin{table}[htbp]
	\centering
	\label{tab1:timer_comparison}
	\caption{Comparison of Timer Management Schemes}
	\begin{tabular}{ c c c c }
		\hline
		\textbf{Source} & \textbf{Implementation} & \textbf{Management} & \textbf{Timing} \\
		& \textbf{Platform} & \textbf{Method} & \textbf{Precision} \\
		\hline
		
		\textbf{jingzhao[10]} & Hardware & One-cycle-check & \si{\micro\second}-level \\
		
		\textbf{AccelTCP[15]} & Software & Timing wheel & \si{\micro\second}-level \\
		
		\textbf{CONNECTX-6[19]}  & Hardware & Unknown & Sub-20 ns \\
		
		\textbf{Proposed} & Hardware & Priority Queue & 12 ns \\
		\hline
	\end{tabular}
\end{table}
scheme to manage these timers, with timer updates implemented via reloading. Modifying a single timer requires 2 cycles, and with a operating frequency of 450 MHz based on 28 nm process, traversing 16K timers takes 72 \si{\micro\second}. ERNIC [11] supports hardware-based automatic packet retransmission and manages hardware timers in the same manner. Scanning all QP timers must be completed within 6554 cycles. Additionally, a fast-clock timer is configured to compensate for timing errors caused by "repeated reloading in a short period," ensuring timing precision. Nevertheless, this scheme faces scalability bottlenecks and limited timing precision, making it suitable for edge RDMA devices with low precision requirements.\\
The ordered list method manages timers by sorting tasks based on their timing values and only checking whether the current value has expired. It can be implemented using either the timing wheel or priority queue.
Taking the timing wheel [17] as an example, it is implemented via a circular array. Each element in the array corresponds to a list, and each list stores tasks with the same expiration time. The current time is represented by a pointer. After each time interval, the pointer moves one unit to check whether there are expired tasks in the list it points to. Typically, the pointer will not jump to the next time unit if elements in the current list have not been fully dequeued. Consequently, the capacity of the list imposes a constraint on timing precision. This method is commonly used in software implementations. The priority queue method works as follows: tasks are sorted in ascending order of their expiration times upon registration. For dequeuing, it only needs to compare the relationship between the first element and the external reference time each time. Moreover, its timing precision is independent of the scale of tasks.\\
The multi-level queue method adopts hierarchical technology to achieve large-range timing, where each level represents a different time granularity and can be implemented using a single timing wheel or a priority queue. However, this scheme consumes substantial resources when adding timers; if the number of timers is large, the system performance will be limited [18].\\
We summarize various existing timer management methods in Table I. Specifically, the timing precision of the Jingzhao RDMA NIC is correlated with the traversal scale and is proportional to the number of timers. In time wheel management implemented via software, the timing precision is generally at the microsecond level—for example, the time interval is set to 100 \si{\micro\second} in [15]. In contrast, our proposed hardware timer queue scheme achieves nanosecond-level timing precision and ensures accurate timing after overflow within a limited bit-width.

In recent years, the academic community has also conducted research on the hardware implementation of priority queues to a certain extent with key application domains including processor task scheduling, packet scheduling, and timer queue management.\\
For task queue-oriented hardware schedulers, [20] implements a priority queue with hardware-software co-management based on a binary heap architecture. However, the authors argue that the drop and decrease\_key operations are used infrequently and are complex to implement in hardware. Thus, only the enqueue and dequeue operations are implemented in hardware, while the drop and decrease\_key operations are handled by software. [21] develops a hardware priority queue named BSRTQ (Basic Shift Register Task Queue) based on shift registers, supporting enqueue, dequeue, and remove operations. It processes enqueue instructions via centralized control and handles other queue control instructions through distributed control. Nevertheless, it does not support hardware-based update operations.\\
For programmable packet scheduling scenarios, [22] proposes a single-instruction-multiple-data (SIMD) priority queue based on a 1D systolic array. In addition to supporting push and pop operations, it adds a replace operation—used to replace the highest-priority element with a new inserted element. [23] presents a tree-based systolic array structure, which also only supports push and pop operations but enables high throughput and large-scale expansion. Owing to the systolic array implementation, the time complexity of each operation is independent of the queue depth.\\
For timer management scenarios, to address the low efficiency of remove operations in timer queues, [24] proposes a priority queue implemented based on systolic arrays, supporting four operations: push, pop, drop, and peek. Each unit is controlled by a state machine. Since the push and pop operations require different numbers of cycles to complete, data conflicts may occur inside the queue, which are resolved by additional control logic. This scheme uses a monotonically increasing timer for timing, which inevitably leads to the problem of external timer overflow—neither this issue nor its solution is mentioned in the paper.
If the solution in [25] is adopted, i.e., resolving timestamp wraparound by increasing the bit-width, a 32-bit counter at a 1 GHz operating frequency wraps around every 4 seconds. When switched to a 64-bit counter, the wraparound interval extends to 584 years, which alleviates the timeout overflow problem to a certain extent. However, from an implementation perspective, the resource overhead nearly doubles.
In contrast, the hardware timer implemented via a d-ary heap in [26] only uses flags to mark overflow exceptions, with no explanation provided on how to handle the overflow after it occurs.
\begin{table*}[!t] 
	\centering
	\caption{Priority Queue function comparsion} 
	\label{tab2:queue_comparison} 
	\begin{tabular}{@{}ccc cccc cc@{}} 
	\toprule 
	\textbf{Priority Queue} & \textbf{Application} & \textbf{Architecture} & \multicolumn{4}{c}{\textbf{Supporting Operations}} & \multicolumn{2}{c}{\textbf{Sorting Method}} \\
	\cmidrule(lr){4-7} \cmidrule(lr){8-9} 
	& & & \textbf{Enqueue} & \textbf{Dequeue} & \textbf{Remove} & \textbf{Update} & \textbf{Global} & \textbf{Group} \\
	\midrule 
	\textbf{BSRTQ}[21] & Task Schedule & 1D Shift Register & $\surd$ & $\surd$ & $\surd$ & $\times$ & $\surd$ & $\times$ \\
	\textbf{BMW-Tree}[23] & Packet Schedule & Tree Systolic Array & $\surd$ & $\surd$ & $\times$ & $\times$ & $\surd$ & $\times$ \\
	\textbf{AnTiQ}[24] & Timer Queue & 1D Systolic Array & $\surd$ & $\surd$ & $\surd$ & $\times$ & $\surd$ & $\times$ \\
	\textbf{Proposed} & Timer Queue & 1D Hybrid & $\surd$ & $\surd$ & $\surd$ & $\surd$ & $\surd$ & $\surd$ \\
	\bottomrule 
\end{tabular}
\end{table*}\\
Hardware priority queues are mostly used in packet schedulers, and they are inherently different from timer queue. Although both sort elements based on priority (timing value), packets become inaccessible from outside the queue once enqueued. Additionally, there is no scenario where the same packet re-enters the queue to modify its priority. Thus, only two basic operations—enqueue and dequeue—need to be supported.
Meanwhile, to handle peak network traffic, packet scheduling takes high throughput and low latency as core requirements. Taking hardware priority queues such as BMW [23] and BBQ [27] as examples, these structures are all designed for work-conserving scheduling [28]: packets are forwarded immediately as long as the link is idle. 
In timer queues, dequeue operation is only executed after the timeout period expires—this aligns with the non-work-conserving scheduling [28]. During the waiting period for high-priority tasks, the link remains idle, with greater emphasis placed on the accuracy of the dequeue timing. Furthermore, timer queues require accessing and modifying timing values of in-queue elements to achieve dynamic updates, making remove a necessary basic operation. 
Table~\ref{tab2:queue_comparison} compares the aforementioned representative hardware priority queue implementations, none of which support update operations. Introducing these technologies into the timer management of priority queues fails to enable priority modification of existing elements or solve the overflow problem caused by fixed bit-width. 

\section{Operations Abstract}
In this section, we first analyze the operation process of basic priority queue operations, and then explain the principles of update operations and grouping sorting.
\subsection{Basic Operations of Priority Queue}
\begin{figure*}[htbp]  
	\centering
	\subfloat[push operation]{
	\includegraphics[width=0.45\textwidth]{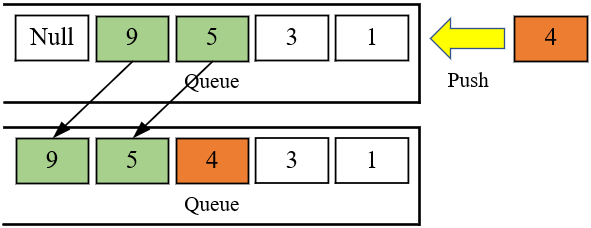}
	}
	\hfill   
	\subfloat[recomve operation]{
	\includegraphics[width=0.45\textwidth]{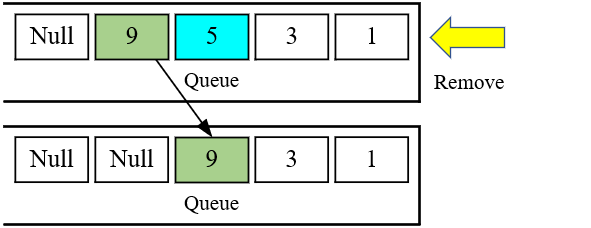}
	}
	\\  
	\subfloat[push\_first operation]{
	\includegraphics[width=0.45\textwidth]{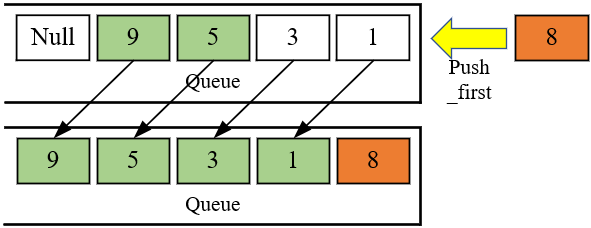}
	}
	\hfill
	\subfloat[pop operation]{
	\includegraphics[width=0.45\textwidth]{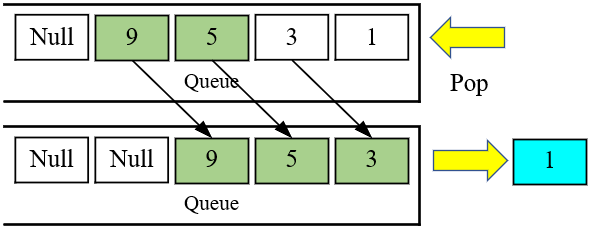}
	}
	\caption{Decomposition of basic operations}
	\label{fig1:basic}
\end{figure*}

In priority queues, elements need to be sorted by their priority, i.e., rank. In practice, each element is also associated with an ID to identify the element. In the design of priority queues for packet schedulers, this ID is also referred to as metadata.
Common priority queues support the following three basic operations:
\begin{itemize}
	\item{Enqueue: adds a new element to the queue;}
	\item{Dequeue: extracts the element with the highest priority from the queue;}
	\item{Remove: eliminates a specific target element from the queue;}
\end{itemize}
From the perspectives of search and translation, enqueue operation is solely associated with searching for the insertion position, while remove operation is only related to searching for the ID. The decomposition of each operation is illustrated in Fig. ~\ref{fig1:basic}. Fig.~\ref{fig1:basic}(a) depicts the push operation: an enqueued element searches for its insertion position by comparing ranks. After identifying the appropriate position, other elements are shifted (i.e., translation followed by setting). Since setting and translation do not conflict at the same position, they can be executed simultaneously.
As shown in Fig.~\ref{fig1:basic}(b), remove operation first involves searching for the target element ID (as indicated by the blue square), followed by a right shift (marked by the green square) to overwrite the target element, thereby completing the deletion process—this is referred to as translation after search.
\\
From the perspective of translation, the pop operation can be considered a special case of remove. Since the element to be deleted is at the queue front, no search is needed—direct shifting to overwrite suffices, as shown in Fig.~\ref{fig1:basic}(d) where the blue block is popped. By analogy, it can be inferred that there must be a special case of the push operation: inserting an element directly at the queue front without searching. We term this operation “push\_first”, illustrated in Fig.~\ref{fig1:basic}(c) where the red block is placed at the front.\\
The four operations mentioned above are referred to as basic operations, and all proceed in two phases: search first, then shifting and positioning. A shifting process is always involved, while the occurrence of search and positioning depends on the specific type of operation. Although the explanation uses a 1D structure as an example, it also applies to tree-based structures—since only one branch of the tree needs to be selected for search. In the existing tree-based structure [23], the branch for element insertion is determined by the counter values along each its path.
\subsection{Update Operation}
Update operation requires both locating the target ID and sorting it according to priority. From the perspective of element translation, we explain the relationship between update operation and basic operations. For generality, we assume an infinitely deep queue composed of several finite-depth subqueues \(q_i\); within a single finite-depth subqueue \(q_i\), the two search methods correspond to four scenarios:
\begin{itemize}
	\item{Locating both the target ID and the priority insertion position;}
	\item{Locating only the target ID, but not the priority insertion position (i.e., the priority of the target element is updated to a lower value);}
	\item{Locating only the priority insertion position, but not the target ID (i.e., the priority of the target element is updated to a higher value);}
	\item{Locating neither the target ID nor the priority insertion position;}
\end{itemize}

Within a queue \(q_i\) of limited depth, locating the target ID corresponds to a remove operation. Due to the continuity of queues, the head element of the adjacent queue \(q_{i+1}\) must be popped, i.e., dequeue operation is performed on \(q_{i+1}\). If the target ID is not found, a remove operation should be propagated to \(q_{i+1}\). Locating the priority insertion position corresponds to a push operation via left shifting and positioning. When elements in \(q_i\) are shifted left, the last element of \(q_i\) is shifted to the front of \(q_{i+1}\) due to queue continuity and the global property that elements are ordered from highest to lowest priority. Thus, when an enqueue operation is valid in \(q_i\), a push\_first operation is propagated to \(q_{i+1}\); if no priority insertion position is found in \(q_i\), an enqueue operation should be propagated to \(q_{i+1}\).When both the ID and the insertion position are found simultaneously, 
\begin{table}[h]
	\centering
	\caption{Operation Propagation Table}
	\label{tab3:operation_propagation}
	\begin{tabular}{|c|c|c|c|c|}
		\hline
		& Enqueue & Remove & Dequeue & Push\_first \\
		\hline
		\begin{tabular}[c]{@{}c@{}}ID Find\\ Rank Find\end{tabular} & $\times$ & $\times$ & $\times$ & $\times$ \\
		\hline
		\begin{tabular}[c]{@{}c@{}}ID Find\\ Rank Not Find\end{tabular} & $\sqrt{}$ & $\times$ & $\sqrt{}$ & $\times$ \\
		\hline
		\begin{tabular}[c]{@{}c@{}}ID Not Find\\ Rank Find\end{tabular} & $\times$ & $\sqrt{}$ & $\times$ & $\sqrt{}$ \\
		\hline
		\begin{tabular}[c]{@{}c@{}}ID Not Find\\ Rank Not Find\end{tabular} & $\sqrt{}$ & $\sqrt{}$ & $\times$ & $\times$ \\
		\hline
	\end{tabular}
\end{table}
meaning that both left and right shifts occur, only partial element displacement happens due to the different shift ranges and no further operations are propagated. Therefore, when searching for both the target ID and priority simultaneously, the above four situations will correspond to the following operations respectively:
\begin{itemize}
	\item{Both remove and enqueue operations are valid, with no propagation to subsequent queues;}
	\item{Only remove operation is valid, propagating dequeue and enqueue operations to subsequent queues;}
	\item{Only enqueue operation is valid, propagating push\_first and remove operations to subsequent queues;}
	\item{Propagating enqueue and remove operation to subsequent queues;}
\end{itemize}
In summary, update operation can be decomposed into a combination of basic enqueue and remove operations, and is completed by propagating these two basic operations. The propagation situations are shown in Table~\ref{tab3:operation_propagation}. Here, "ID Find" means locating the ID in the current queue, and "Rank Find" refers to finding the appropriate insertion position in the current queue. Therefore, update operation can be realized by implementing the combination and propagation of these basic operations.
\subsection{Group Sorting}
During the operation of a priority queue, if elements with higher priorities are continuously enqueued and dequeued, low-priority elements within the queue will never be dequeued—a phenomenon known as starvation. This occurs because the search process adheres to the principle that priorities are ordered from highest to lowest at any position.
Precisely this property makes it impossible to ensure timing accuracy after timer overflow.
If the queue is divided into groups, where the descending priority order is enfored intra-group (rather than globally), while the ordering between groups remains flexible. This arrangement permits the dequeuing of lower-priority elements even when higher-priority elements are present, effectively introducing a latency for the latter. This mechanism is termed group sorting.\\
Group sorting is defined as follows: Elements in the queue are divided into two groups based on a boundary priority. Elements within each group remain sorted in descending order of priority, while the order between groups is variable. During enqueue, elements are searched for and inserted according to the inter-group order. Using notations from Order Theory, the rules for priority queues supporting group sorting are defined as follows:
\begin{table}[h]
	\centering
	\caption{Enqueue sorting Rule}
	\label{tab4:enqueue_rule}
	\begin{tabular}{|c|c|c|c|}
		\hline
		Head vs Boundary & Sorting Rule & $P_e$ vs $P_l$ & Position \\
		\hline
		\multirow{2}{*}{$P_h \succ P_l$} & \multirow{2}{*}{$\{Q2, Q1\}$} & $P_e \succ P_l$ & Fig. 2(a) \\
		\cline{3-4}
		&  & $P_e \preceq P_l$ & Fig. 2(b) \\
		\hline
		\multirow{2}{*}{$P_h \preceq P_l$} & \multirow{2}{*}{$\{Q1, Q2\}$} & $P_e \succ P_l$ & Fig. 2(c) \\
		\cline{3-4}
		&  & $P_e \preceq P_l$ & Fig. 2(d) \\
		\hline
	\end{tabular}
\end{table}
\begin{itemize}
	\item{The priority range of queue Q is \(R_Q = [P_{\text{min}}, P_{\text{max}}]\), where \(P_{\text{max}}\) denotes the highest priority and \(P_{\text{min}}\) denotes the lowest priority. Define a boundary priority \(P_l\) satisfying \(P_l \in (P_{\text{min}}, P_{\text{max}})\). The priority interval of subgroup \(Q_1\) is \([P_l, P_{\text{max}}]\), and that of subgroup \(Q_2\) is \([P_{\text{min}}, P_l)\). Elements within each subgroup are sorted in descending order of priority (from highest to lowest). Two possible orders exist between \(Q_1\) and \(Q_2\): \(\{Q_1, Q_2\}\) or \(\{Q_2, Q_1\}\);}
	
	\item{When the priority \(P_h\) of \(\text{head}(Q)\) is higher than \(P_l\) (i.e., \(P_h \succ P_l\)), the enqueuing sorting rule follows \(\{Q_2, Q_1\}\);}
	
	\item{When the priority \(P_h\) of \(\text{head}(Q)\) is not higher than \(P_l\) (i.e., \(P_h \preceq P_l\)), the enqueuing sorting rule follows \(\{Q_1, Q_2\}\);}
\end{itemize}
According to the above rules, the relationship between subgroups is determined by the comparison between the head priority \(P_h\) and the boundary priority \(P_l\), while elements within each subgroup remain sorted in descending order of priority. For an enqueuing element \(e\), finding its insertion position involves two phases: first, determining which subgroup it belongs to; second, comparing its priority with others within that subgroup to locate the insertion spot. Based on the above definitions, if the priority of the enqueuing element \(e\) is \(P_e\), the enqueuing sorting rules are summarized in Table~\ref{tab4:enqueue_rule}.\\
Fig. 2(a) and 2(b) illustrate the sorting rule where the entire queue maintains elements in descending order of priority. As high-priority elements are dequeued sequentially until \(P_h \preceq P_l\), the lower-priority group Q2 is dequeued first, with no remaining Q1 elements in the priority queue at this point. When an enqueued element \(e\) belongs to Q1 (i.e., \(P_e \succ P_l\)), it skips comparisons with all elements in Q2 according to the sorting rules, until the first empty slot or Q1 is found. Dequeuing of Q1 elements is allowed only after all elements in Q2 have been dequeued. Thus, group sorting adjusts the element order by altering insertion positions. In cases of timer overflow (i.e., priority elevation), the sorting rule in Fig. 2(c) enables delayed dequeuing of overflow values, thereby resolving the overflow issue.
\begin{figure*}[htbp]  
	\centering
	\subfloat[Sorting rule when $P_h \succ P_l$,$P_e \succ P_l$]{
		\includegraphics[width=0.45\textwidth]{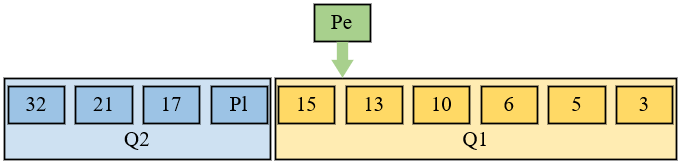}
	}
	\hfill   
	\subfloat[Sorting rule when $P_h \succ P_l$,$P_e \preceq P_l$]{
		\includegraphics[width=0.45\textwidth]{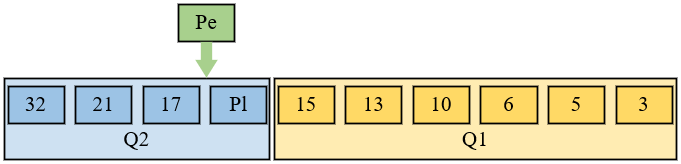}
	}
	\\  
	\subfloat[Sorting rule when $P_h \preceq P_l$,$P_e \succ P_l$]{
		\includegraphics[width=0.45\textwidth]{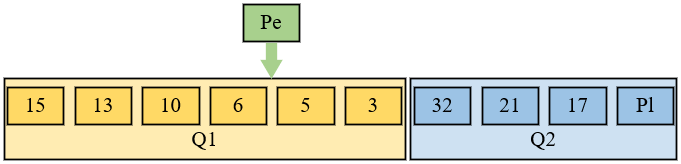}
	}
	\hfill
	\subfloat[Sorting rule when $P_h \preceq P_l$,$P_e \preceq P_l$]{
		\includegraphics[width=0.45\textwidth]{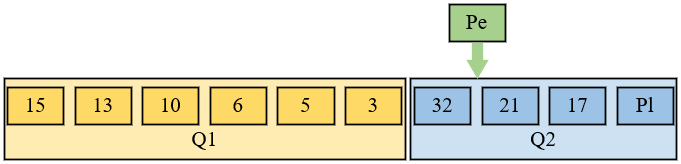}
	}
	\label{fig2:group}
	\caption{Insertion Position by Group Sorting Rule}
\end{figure*}
\section{Architecture Design}
In this section, we first design the overall architecture of the priority queue, along with systolic units and its timing sequence. Finally, we design a comparator structure to realize the group sorting.
\subsection{Hybrid Architecture}
Common implementation schemes for hardware priority queues include 1D structures and tree-based structures. Tree-based structures offer high scalability but are unfavorable for operation propagation due to their numerous branches. Among one-dimensional structures, 1D systolic arrays operate at a higher frequency than shift registers; however, shift registers incur lower resource overhead compared to systolic arrays [30]. Thus, we adopt a hybrid structure of 1D systolic arrays and shift registers as the final implementation. 
To reduce repeated element movement when combining and propagating basic operations, we expand the comparison range within a single systolic unit. Additionally, we optimize the implementation method for maintaining the first-in-first-out (FIFO) order of elements with the same priority in the systolic array.
\begin{figure*}  
	\centering
	\includegraphics[width=1\textwidth]{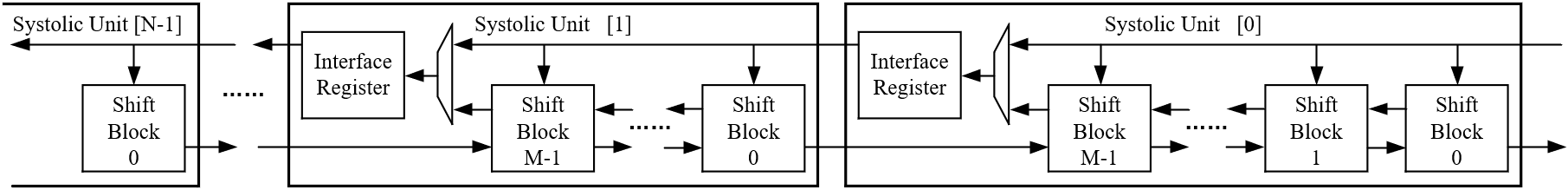}
	\label{fig3:Hybrid}
	\caption{Hybrid Architecture}
\end{figure*}\\
The hybrid structure, as shown in Fig. 3 (a simplified structural diagram under the push operation only), mainly consists of two modules: the Systolic Unit and the Shift Block. A systolic array is formed by cascading N Systolic Units, with interface\_register model between each pair of systolic units to propagate operations to the next stage. Each Systolic Unit is composed of M cascaded Shift Blocks, allowing the queue to have a maximum capacity of \(N*M\) elements. Each element within a Shift Block contains two fields: ID and DATA, whose meanings are as follows: 
\begin{itemize}
	\item{ID: This field corresponds to the identifier of each element, which is unique within the queue. It is initialized to all zeros; zero does not represent any element and indicates that the current position is empty;}
	\item{DATA: Stores the timing value corresponding to each element, and sorting is performed based on the comparison results of these values. It is initialized to all ones;}
\end{itemize}
The internal structure of the Systolic Unit is shown in Fig. 4, where dashed lines
\begin{figure}[htbp]
	\centering  
	\includegraphics[width=0.5\textwidth]{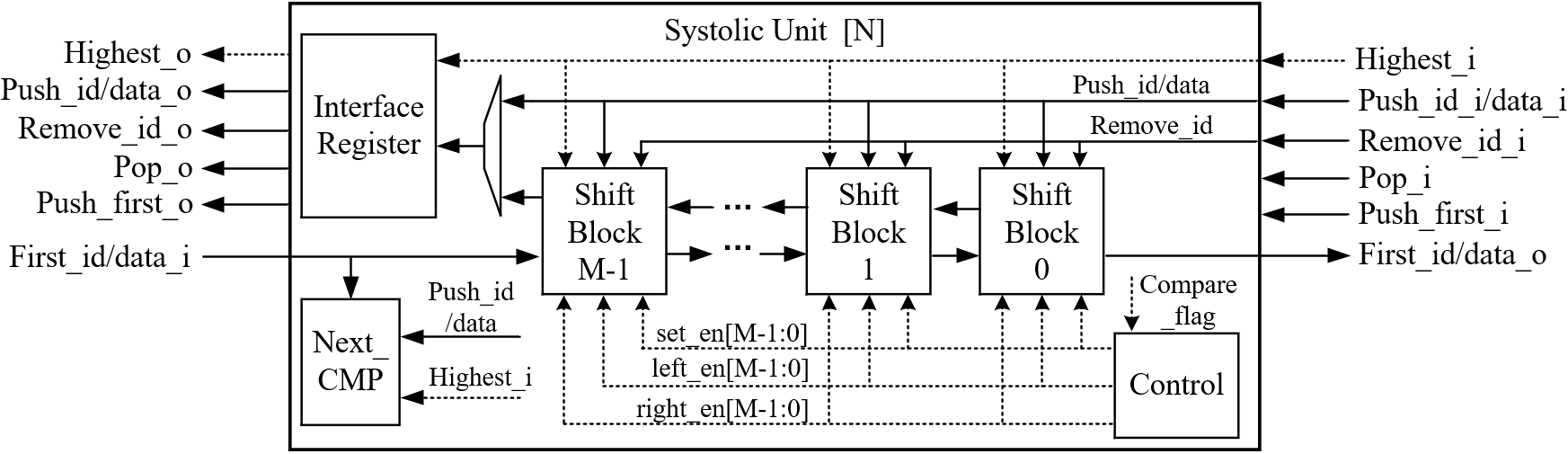}  
	\caption{Systolic Unit}  
	\label{fig:systolic}  
\end{figure}
represent control logic and solid lines indicate element transmission. Elements are sorted from right to left in the queue based on increasing DATA values (i.e., descending priority). While update and enqueue operations differ in that an enqueue inserts a new element whereas an update modifies an existing one, they are treated identically at the implementation level: both are executed as push operations via the \(Push\_id\_i\) and \(Push\_data\_i\) ports.\\
Upon entering a systolic unit, an element’s priority is compared against a total of M+1 elements: the M resident elements of the current unit and the head element of the next-level unit, identified by \(First\_id\_i\) and \(First\_data\_i\)). This comparison is executed by the Next\_CMP model and the comparators within each shift block, as depicted in Fig. 4. This design preemptively resolves a potential inefficiency where the concurrent propagation of two valid operations could cause redundant movement of the next-level’s head element, should a new element be inserted into the current unit’s tail position. By directly comparing with the next-level head, this mechanism significantly reduces cycle overhead.\\
The result of all these comparisons is aggregated into a signal named \(compare\_flag\)) in Fig. 4. Given that the ID search result is a one-hot code and the DATA comparison yields a binary sequence of consecutive 0s and 1s, Boolean logic operations can be used to derive the control signals for each Shift Block. 
Furthermore, the \(Push\_first\_i\) signal of each systolic unit implements the push\_first operation, which preserves the FIFO order for elements of identical priority. This issue of maintaining FIFO order is addressed in [30] by augmenting each element with a dedicated flag bit.\\
Fig. 5 presents a timing diagram illustrating the propagation of operations across multiple systolic units, where each operation incurs an overhead of 3 cycles. When two operations are valid simultaneously, the combinational logic path for resolving the shift and set signals via Boolean logic becomes relatively long. To mitigate this, the search phase is decoupled from the shift-and-set phase and pipelined across two separate cycles. 
In the finish phase, registers complete the positioning; at this point, the next-level head element stabilizes, allowing comparison and search to proceed. Simultaneously, operations are propagated to the subsequent Systolic Unit.\\
In principle, pop and push\_first operations do not require a comparison process. However, since two types of operations need to be propagated simultaneously, they are still designed to complete within 3 cycles. This design choice is made to prevent potential hazards that arise from operations completing in different cycle counts when both must be propagated concurrently.
\begin{figure}  
	\centering
	\includegraphics[width=0.5\textwidth]{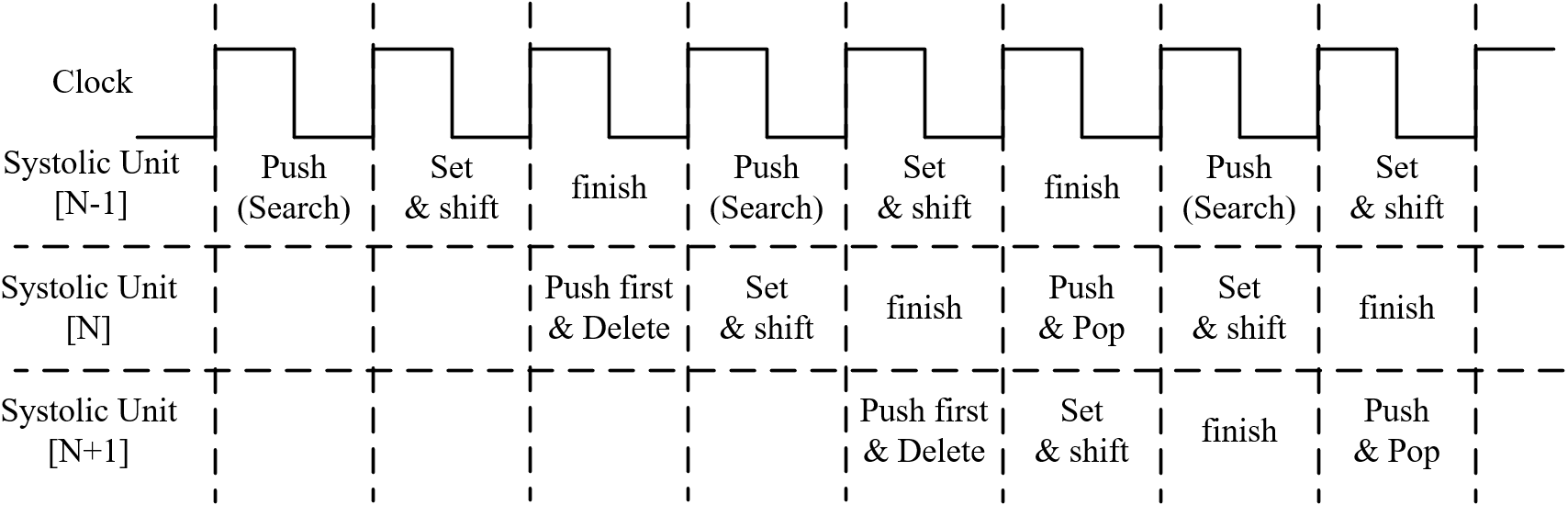}
	\label{fig:timing}
	\caption{Timing Graph of Propagating Operations}
\end{figure}
\subsection{Implementation of Group Sorting}
In timer queue management based on a priority queue, one method for calculating expiration time involves adding a fixed timeout to an external reference timer. Specifically, the DATA calculation is shown in equation (1), where \(R_t\) denotes the external reference timer with a bit-width of \(W_r\)—consistent with the bit-width of DATA. \(R_t\) increments at a set timing precision and has a timing range of \([{\text{0}},2^{W_r} - 1]\), which corresponds to priorities from \(P_{max}\) (highest) to \(P_{min}\) (lowest). \(TO\) represents the timeout set for each task, with a bit-width of \(W_o\); thus, the configurable timeout range is \(({\text{0}},2^{W_o} - 1]\), where \(W_r > W_o\).
\begin{align}
	DATA &= R_t+TO
\end{align}
After addition according to equation (1), the bit-width of DATA may extend to \(W_r + 1\), which means an overflow occurs. Moreover, as time elapses, \(R_t\) will also overflow. Let \(P_e\) denote the DATA value of an element \(e\) to be enqueued. When sorting by global priority in descending order, if DATA overflows while \(R_t\) does not, \(P_e\) must be less than \(R_t\). In this case, the overflowed element will be dequeued immediately, iolating its intended timeout period \(TO\). Here, the overflow value is recorded as
\begin{figure}[htbp]
	\centering  
	\includegraphics[width=0.5\textwidth]{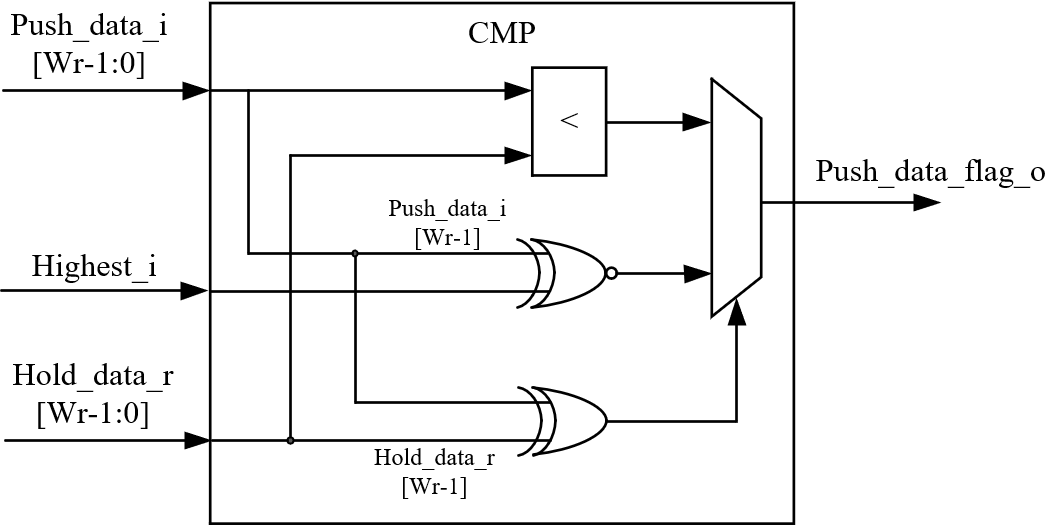}  
	\caption{Overflow Control Comparator}  
	\label{fig:cmp}  
\end{figure}
\(P_e' = TO - (2^{W_r} - R_t')\). However, the overflowed element should expire only after \(TO\) time units. With group sorting, as shown in Fig. 2(c), after \(2^{W_r} - R_t'\) time units, all elements in group Q2 in Fig. 2(c) will be dequeued, and the queue will return to the state shown in Fig 2(a). Then, \(P_e\) is dequeued after another \(TO - (2^{W_r} - R_t')\) time units. Without updating \(P_e\) the total duration from its enqueue to dequeue is exactly \(TO\) time units. Thus, when DATA overflows, the sorting method in Fig 2(c) enables correct dequeuing timing. Whether or not overflow occurs in DATA and \(R_t\), proper dequeuing timing is achieved using the methods shown in Fig. 2(a), 2(b), or 2(d). This indicates that group sorting can resolve timing errors caused by overflow.\\
According to the group sorting rules, when implementing group sorting in the timer queue, a boundary value \(P_l\) needs to be set. According to the principle of Fig. 2(c), \(P_e\) does not need to be compared with any elements in Q2; it only undergoes comparison and insertion after entering the range of Q1. Thus, the insertion comparison logic must be adjusted. In this design, to reduce resource overhead and leveraging the characteristics of binary systems, we choose to evenly divide the priority range, i.e., \(P_l = 2^{W_r-1}\). Thus, the range of Q1 is \([0, 2^{W_r-1} - 1)\) and the range of Q2 is \([2^{W_r-1}, 2^{W_r} - 1]\). At this time, it is required that $W_r > W_o + 1$. 
In implementation, the distinction between group Q1 and Q2 can be made by comparing only the MSB of the incoming element’s DATA with that of the elements in the queue. If the MSBs match, a full priority comparison based on the DATA value is performed within the group. If the MSBs differ, the comparison in the current group is skipped. This mechanism efficiently fulfills the requirements of group sorting. 
Compared to significantly increasing the bit width, this scheme effectively reduces resource overhead.\\
Fig. 6 shows the implementation scheme of the overflow control comparator, which is used to compare the timing values in each shift block and also corresponds to the implementation of Next\_CMP in Fig. 4. Here, \(Highest\_i\) denotes the MSB of the head element’s DATA, matching the \(Highest\_i\) signal in Fig. 4. When \(HIGH\), this signal indicates that the head element’s priority is greater than or equal to \(P_l\). Meanwhile, \(Hold\_data\_r\) is a shift register signal that stores the elements. \(Push\_data\_flag\_o \) is the comparison result and is used for Boolean logic operations.
\section{Use Case}
To demonstrate the effectiveness of updates and group sorting, we applied the hardware timer queue to manage the timeout of entries in a flow table. Simulations were conducted by configuring two timing precisions and multiple timeout values. The external clock\begin{table}[htbp]
	\centering
	\caption{Different DATA Width \(W_r\)}
	\label{tab5}
	\begin{tabular}{|r|c|c|c|c|c|}
		\hline
		Timer Width & \(p\) & \( TO\) & Pop & Max & Mpps\\
		\(W_r\) & /Period & /Period & number & Queue Occ & \\ 
		\hline
		9, 10, 11, 12 & 6 & 127  & 14821 & 107 & 166.41 \\
		\hline
		10, 11, 12 & 6 & 191  & 12635 & 127 & 166.26 \\
		\hline
		10, 11, 12 & 6 & 255  & 11733 & 147 & 166.12 \\
		\hline
		10, 11, 12 & 6 & 383  & 10649 & 185 & 165.81 \\
		\hline
		11, 12 & 6 & 511  & 8938  & 215 & 165.52 \\
		\hline
		12 & 6 & 1023 & 6708  & 321 & 164.27 \\
		\hline
	\end{tabular}
\end{table}reference timer increments continuously: when a data packet arrives, the new expiration time for the corresponding flow is calculated and enqueued; when the time of the head element is less than the external reference timer, a pop operation is executed. The data packets used a subset of the UNIV1 dataset from [31]. Flows with identical TCP 5-tuples in the dataset were counted as stimuli. All experiments tested 2047 flows and a total of 119,870 data packets, with the queue depth set to 2048 and the clock cycle to 2 \(ns\)—resulting in a theoretical queue throughput of 166.6 Mpps. Since the maximum utilization of the simulated queue in all experiments did not exceed 50\%, no remove operation was added to the simulations; only push and pop operations were included.\\
A single operation takes 3 cycles to complete, and one pop operation dequeues only one task. There may be many tasks with the same expiration time in the queue; in such cases, an alternating push-and-pop operation method is adopted to reduce the waiting time for dequeuing tasks with identical expiration times, as well as the waiting time for external update or enqueue operations. To ensure that at least one alternating push-and-pop cycle is completed within a single timing precision interval, the timing precision \(p\) is set to 6, meaning the external reference timer increments by 1 every 6 clock cycles (12 \(ns\)).\\
The experimental results for various timer widths \(W_r\) and timeout values \(TO\) are presented in Table~\ref{tab5}. It can be observed that for a given \(p\) and \(TO\), the simulation results are identical across all tested timer widths. This indicates that under the condition of identical \(p\) and \(TO\) settings, all simulation results are independent of the timer width \(W_r\).The reason is that the number of flow timeout updates is tied to the packet arrival frequency: each arriving packet triggers an update to the corresponding flow’s expiration time. This update process requires no software intervention in queue control, being solely driven by incoming packets. This eliminates the need for timer checks and flow arrival statistics in software algorithms, which helps reduce the software load on the control plane. Additionally, group sorting ensures correct ordering and timing after expiration time overflow. Thus, with the same packet arrival frequency, the number of updates and the increment of expiration time remain consistent, leading to identical simulation results. When \(TO\) is set to 127, the operation throughput reaches 166.41 Mpps, approaching the theoretical throughput—indicating that the system operates near a work-conserving schedule. As the timeout increases, the number of pop operations decreases, and the operation throughput declines, 
\begin{table}[h]
	\centering
	\caption{Timer Precision \(p\)=12}
	\label{tab6}
	\begin{tabular}{|c|c|c|c|c|c|}
		\hline
		Timer Width & \(p\) & \( TO\) & Pop & Max & Mpps\\
		\(W_r\) & /Period & /Period & number & Queue Occ & \\ 
		\hline
		12 & 12 & 127 & 11742 & 145 & 166.12 \\
		\hline
		12 & 12 & 191 & 10674 & 184 & 165.81  \\
		\hline
		12 & 12 & 255 & 8949  & 214 & 164.52  \\
		\hline
		12 & 12 & 383 & 7600  & 271 & 164.92 \\
		\hline
		12 & 12 & 511 & 6716  & 321 & 164.28 \\
		\hline
		12 & 12 & 1023 & 4406 & 496 & 161.66 \\
		\hline
	\end{tabular}
\end{table} 
\begin{figure}[htbp]
	\centering  
	\includegraphics[width=0.48\textwidth]{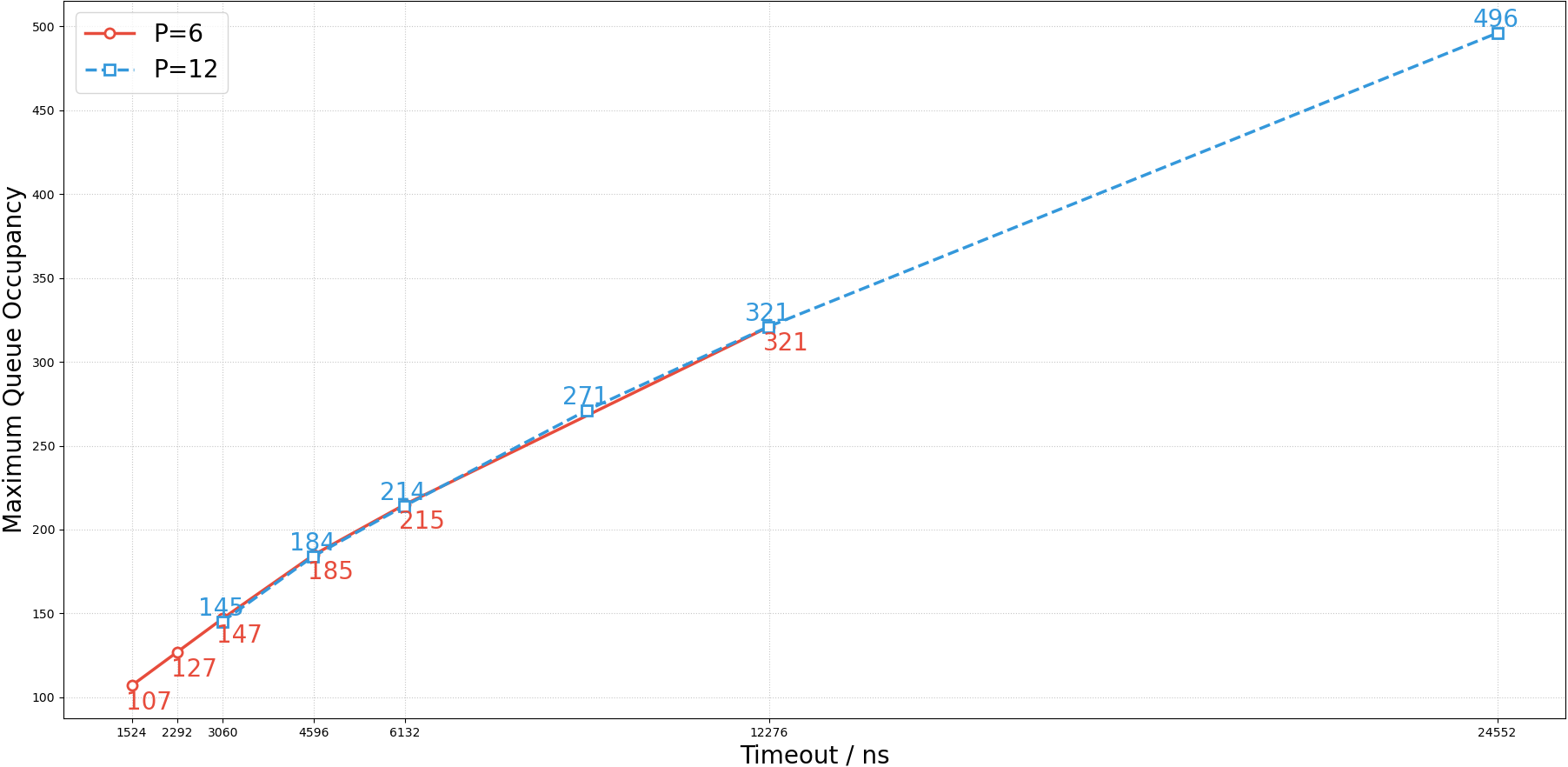}  
	\caption{Maximum Queue Occupancy of Different Precision}  
	\label{fig7:depth}  
\end{figure}
further confirming operation in a non-work-conserving schedule mode.\\
In the same scenario, we tested a timer queue that does not support group sorting. For the experiment in Table~\ref{tab5} with a maximum queue occupancy of 107 (i.e., timing precision \(p = 6\), timeout = 127), we modified our proposed scheme by increasing the bit-width to match the requirements. Both methods yielded identical simulation results over a total duration of 809,383 \(ns\). However, the timer queue using the bit-width extension approach required a timer width \(W_r\) of 17 bits, whereas the timer queue supporting group sorting only needed a \(W_r\) of 9 bits.If the simulation time were extended, expiration times would increase further. The timer width of the non-group-sorting queue would eventually exceed 17 bits, while the group-sorting-enabled queue maintains correct sorting with a fixed 9-bit width that does not scale with simulation time. Thus, group sorting effectively ensures the accuracy of dequeuing timing after overflow with significantly lower resource overhead.\\
Additionally, the timing precision \(p\) was set to 12. The simulation results are presented in Table~\ref{tab6}. A comparison between Table~\ref{tab5} and Table~\ref{tab6} reveals the following: as timing precision increases, the number of pop operations decreases and the maximum queue occupancy increases—this phenomenon is consistent with that observed when only the timeout is increased. The reason for this is that increasing either timing precision or timeout prolongs the duration for which timers remain in the queue. This, in turn, increases the probability of updates and reduces the number of pop operations. Another observation from comparing Table~\ref{tab5} and Table~\ref{tab6} is that when \(p \times TO\) values are close, the simulation results are also close. Taking the maximum queue occupancy as an example (as shown in Fig. 7), the depth curve for timing precision \(p = 6 \) and that for \(p = 12 \) are highly overlapping and follow the same trend. This occurs because the packet stimuli are identical, and when \(p \times TO\) values are close, the total duration for which flows remain in the queue is similar—leading to comparable simulation results. Thus, to maintain the same effective timeout, modifying only the timing precision or only increasing the timeout can achieve the same goal. Therefore, this timer queue design can meet the requirements of high-precision timing, scale expansion, and dynamic hardware updates.

\section{Evaluation}
\subsection{Prototype Implementation}
The hardware priority queue we proposed is written in Verilog and verified using SystemVerilog. To evaluate its performance and resource overhead, we conducted assessments based on FPGA and a 28nm process respectively. The FPGA synthesis and implementation were carried out using Xilinx Vivado 2024.2, with the Xilinx VCU118 board as the platform, which has 1182K LUTs and 2364K flip flops. The reconfigurable parameters of the queue include: ID width, DATA width, the number of Systolic units (N), and the number of Shift Blocks (M) in each Systolic unit—with the requirement that \(M \geq 2\). 
By setting different values of N and M, we studied the patterns of clock frequency and resource overhead under the same queue depth. In addition, to demonstrate that our design effectively resolves the overflow issue of hardware timers with fixed bit-width, we compared its resource overhead with that of a structure that does not support overflow when implementing the same simulation scenario. Finally, we also synthesized the open-source timer queue AnTiQ [24] for comparison purposes.
\subsection{Cost and Performance}
The queue depth is fixed at 4K, supporting a 16-bit timer (i.e., 64K priorities). Different values of N and M are configured to investigate the performance and resource overhead of various structures under the same queue capacity. Since the queue depth is 4K (i.e., the queue can hold a maximum of 4096 elements), the ID width is set to 12 bits. The resource overhead and throughput of different structures are shown in Fig. 8. With the same queue depth and number of priorities, as M increases, the downward trend of LUTs and FFs slows down and even stabilizes. When \(M = 2\), \(N = 2024\) accordingly, and the area reaches its maximum. This structure has the smallest combinational logic delay, so theoretically, the throughput should be maximized. However, due to the presence of routing delays in FPGAs—which account for the majority of total delay—the maximum throughput of 116 Mpps is achieved when \(M = 4\). As M increases and N decreases (with the total number of Shift Blocks remaining constant), the number of systolic units decreases, leading to a rapid reduction in the total number of LUTs and FFs. 
\begin{figure}[htbp]  
	\centering
	\subfloat[LUTs and FFs Consumption]{
		\includegraphics[width=0.45\textwidth]{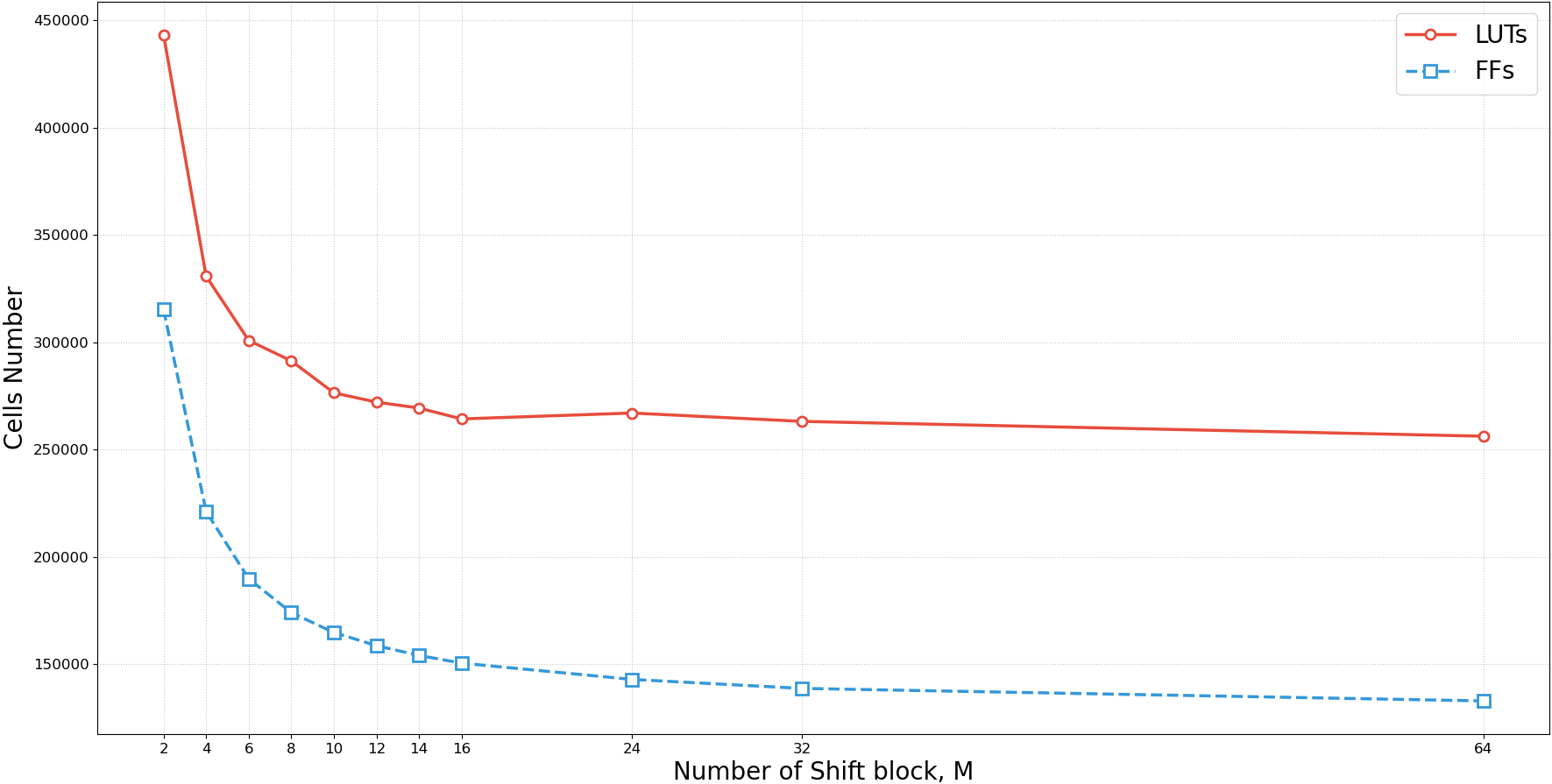}
	}
	\hfill   
	\subfloat[Maximum Frequency]{
		\includegraphics[width=0.45\textwidth]{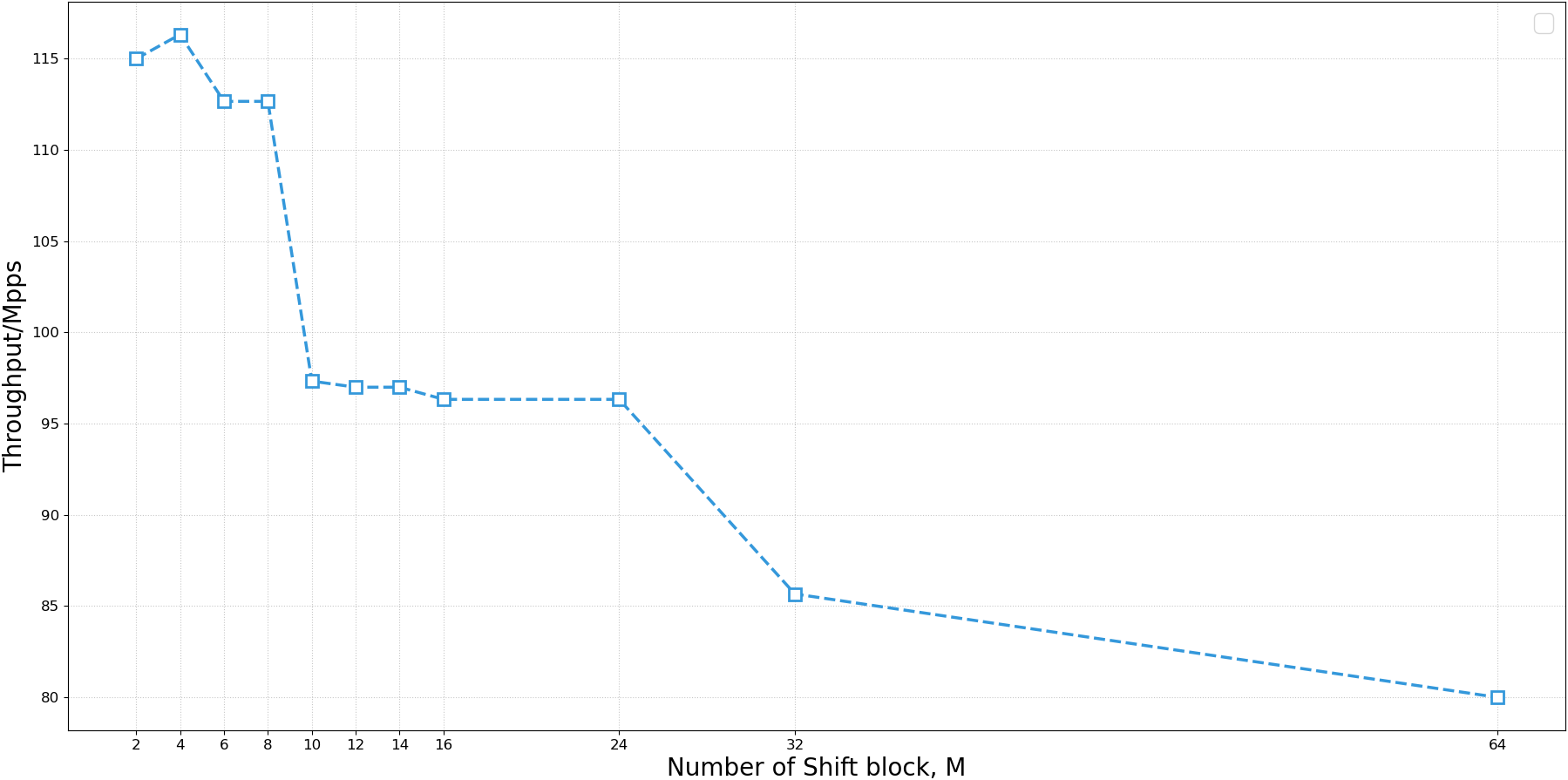}
	}
	\label{fig8:fpga}
	\caption{FPGA Implemetation Results}
\end{figure}
Throughput decreases rapidly, which is because the bus load increases with M during broadcast comparisons within a single systolic unit, increasing combinational logic delay. The increased broadcast load also exacerbates routing delays after placement in the FPGA. The combined effect of increased bus load and routing delays reduces the operating frequency on the FPGA, thereby lowering throughput.\\
As a register-based priority queue supporting a capacity of 4096, PIFO [28] completes a single operation in just 1 cycle but achieves a throughput of only 40 Mpps. In contrast, our proposed architecture requires 3 cycles per operation, yet it delivers a throughput 2.9 times higher. This advantage is mainly attributed to the 1D systolic array, which reduces bus load and thereby shortens the critical path delay. Conversely, R-BMW [23], which also supports a 4K capacity, achieves a throughput of 192 Mpps (1.6× our design) primarily because it completes an operation in only 2 cycles. Our design’s longer operation cycle stems from the need to handle update operations, which can involve two basic operations occurring concurrently. To manage this complexity and reduce combinational logic delay, we deliberately partition the comparison and positioning phases into two separate cycles, leading to a 3-cycle operation. As a result, our throughput is 60\% of that of R-BMW.
\begin{figure*}[htbp]  
	\centering
	\subfloat[Synthesis Area]{
		\includegraphics[width=0.45\textwidth]{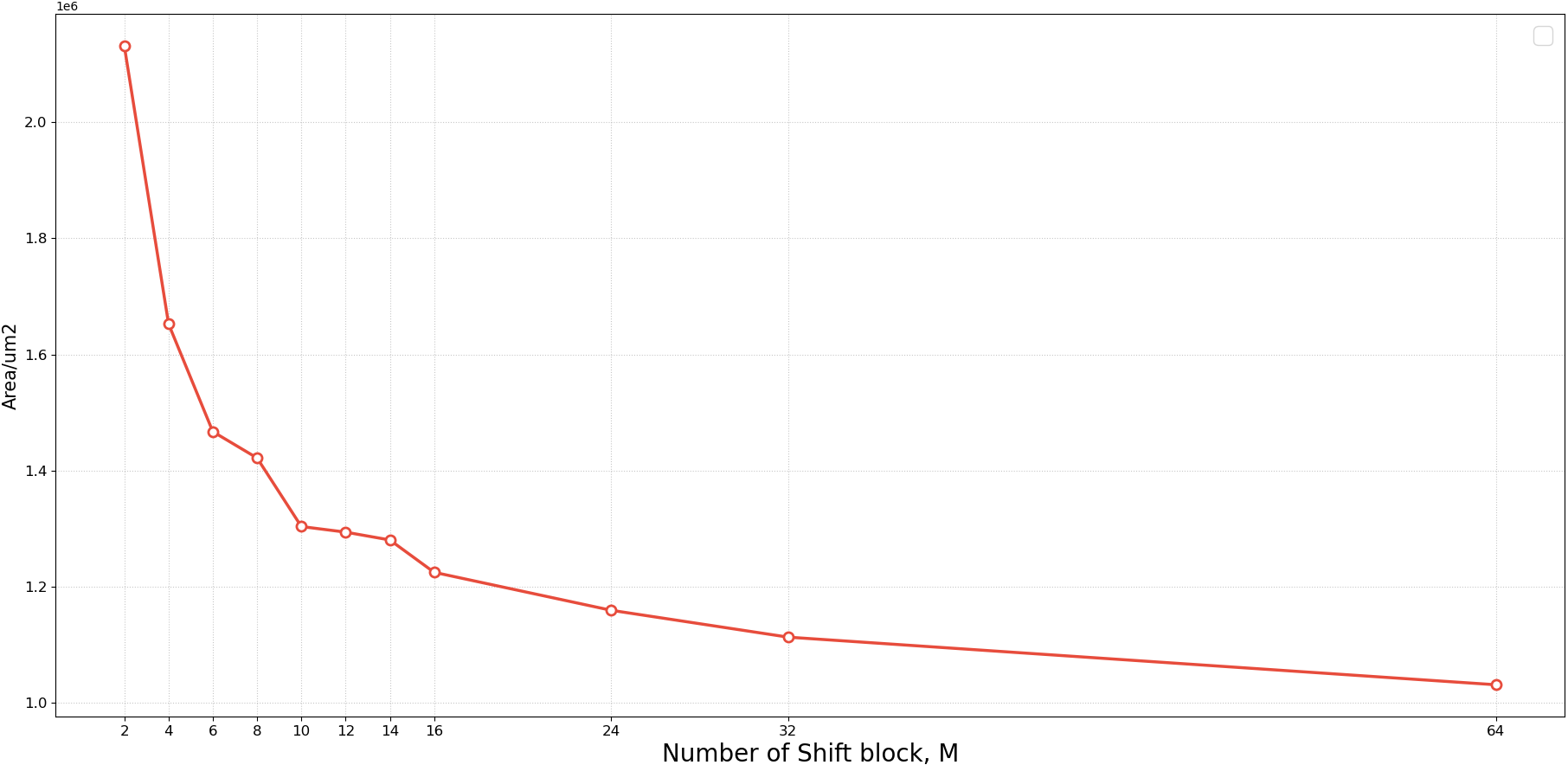}
	}
	\hfill   
	\subfloat[Maximum Frequency]{
		\includegraphics[width=0.45\textwidth]{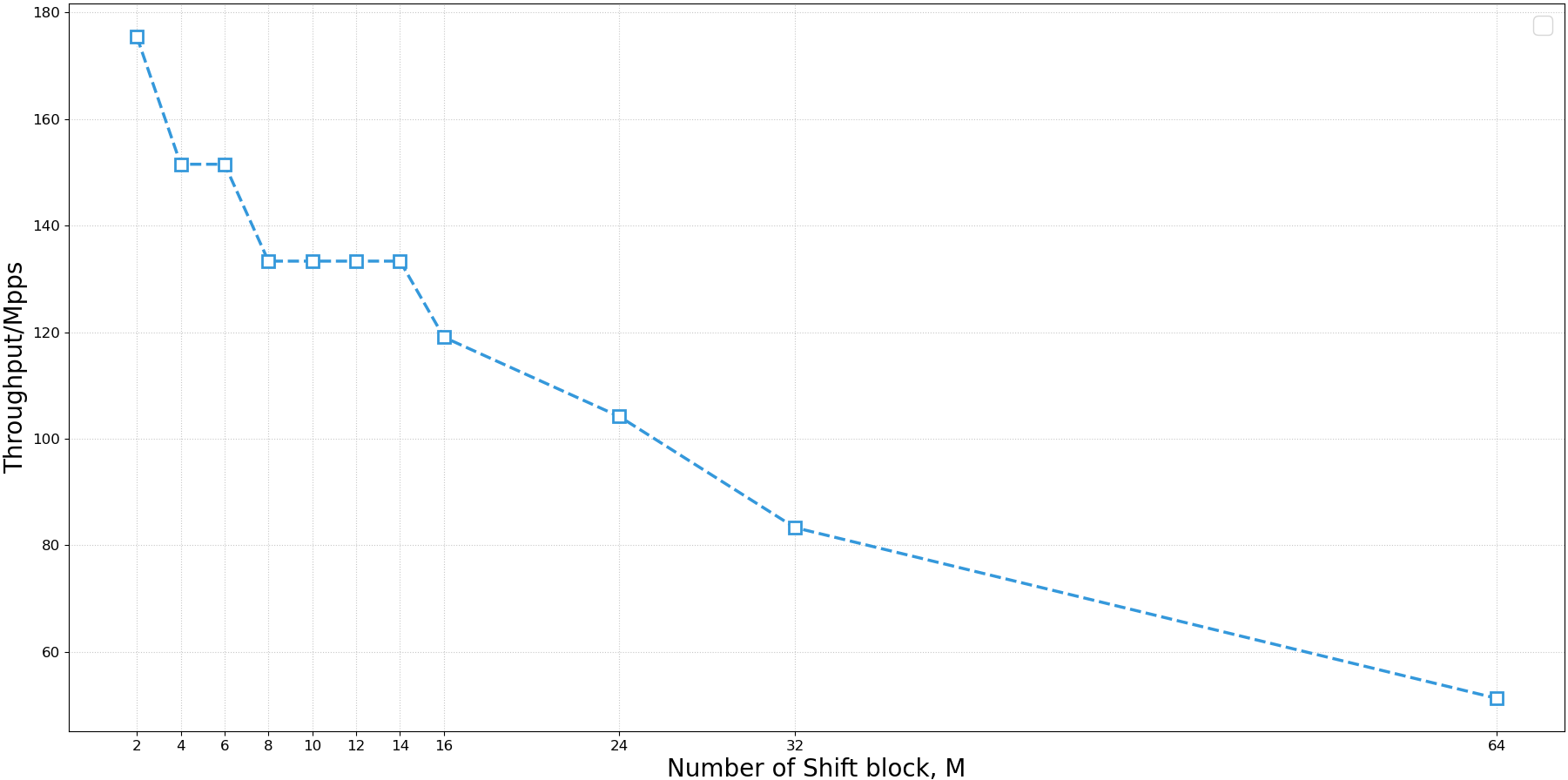}
	}
	\label{fig9:ASIC}
	\caption{28nm Synthesis Results}
\end{figure*}\\
The architectures, based on the aforementioned parameters, were synthesized using a 28nm process. Fig. 9 presents the resulting area and throughput. For a constant queue depth, a trend of decreasing total area is observed as M increases and N decreases, although the rate of this reduction gradually diminishes. Conversely, increasing M elevates the bus load, which in turn extends the critical path delay and degrades throughput. The maximum throughput of 175 Mpps is achieved at M=2, operating at 526 MHz with a critical path delay of 1.82 \(ns\). This satisfies the 2 \(ns\) cycle time requirement from the packet simulation, which corresponds to a timing precision of 12 \(ns\). An analysis of the Area-Delay Product (ADP) reveals that the minimum ADP for a 4K queue depth is attained at M=14, yielding a throughput of 133 Mpps. The second-best ADP is achieved at M=6, with a corresponding throughput of 151 Mpps. For comparison, a PIFO supporting the same queue depth, when synthesized, exhibits a critical path delay of 16.93 \(ns\) and an area of approximately 1,125,966 ~\si{\um^2}. While its area is comparable to our design at M=24 and N=171, our proposed architecture achieves a critical path delay of only 2.7 \(ns\), representing an 84\% reduction in latency.
\begin{figure*}[htbp]  
	\centering
	\subfloat[LUTs and FFs Consumption]{
		\includegraphics[width=0.45\textwidth]{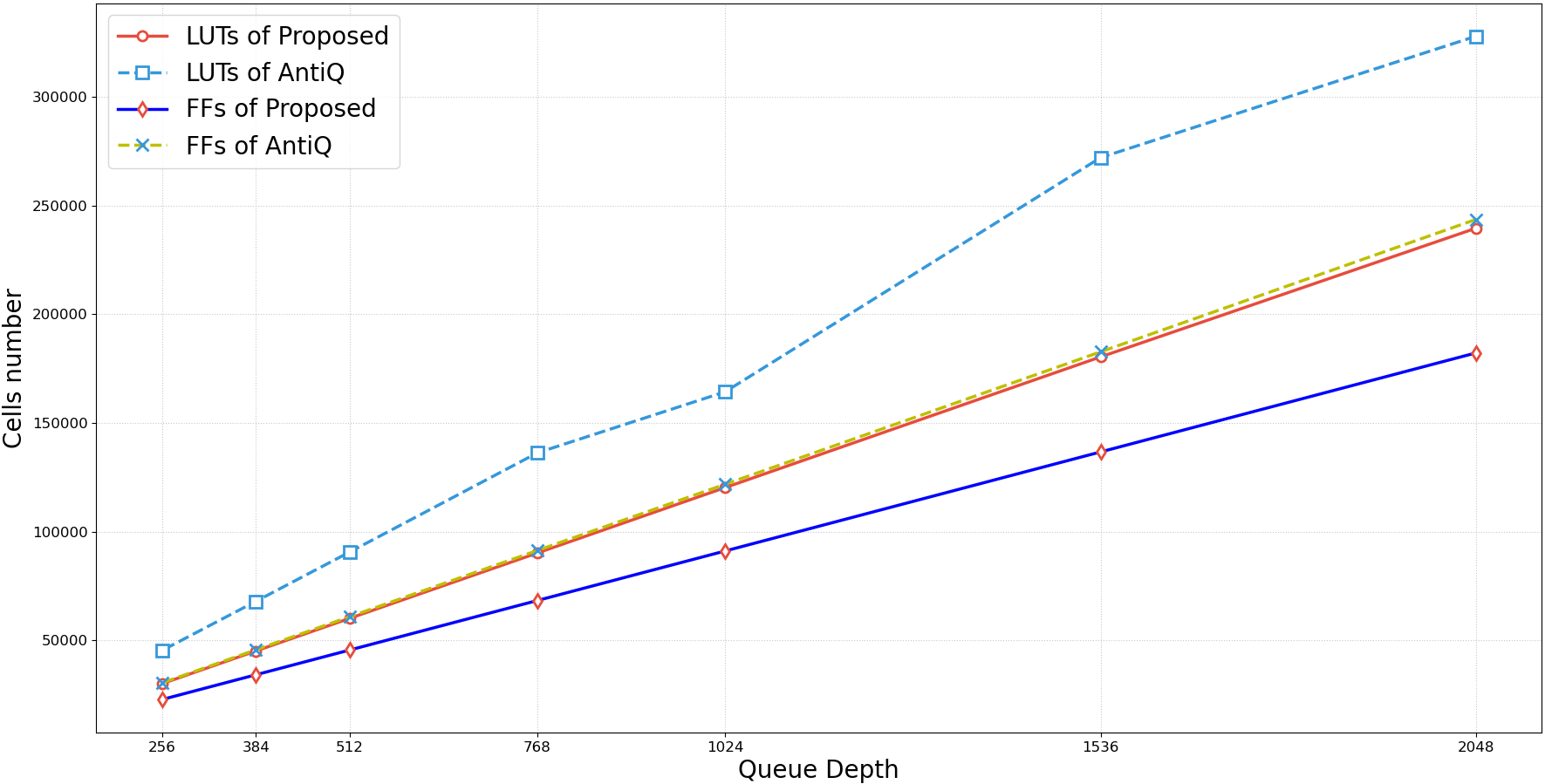}
	}
	\hfill   
	\subfloat[Maximum Frequency]{
		\includegraphics[width=0.45\textwidth]{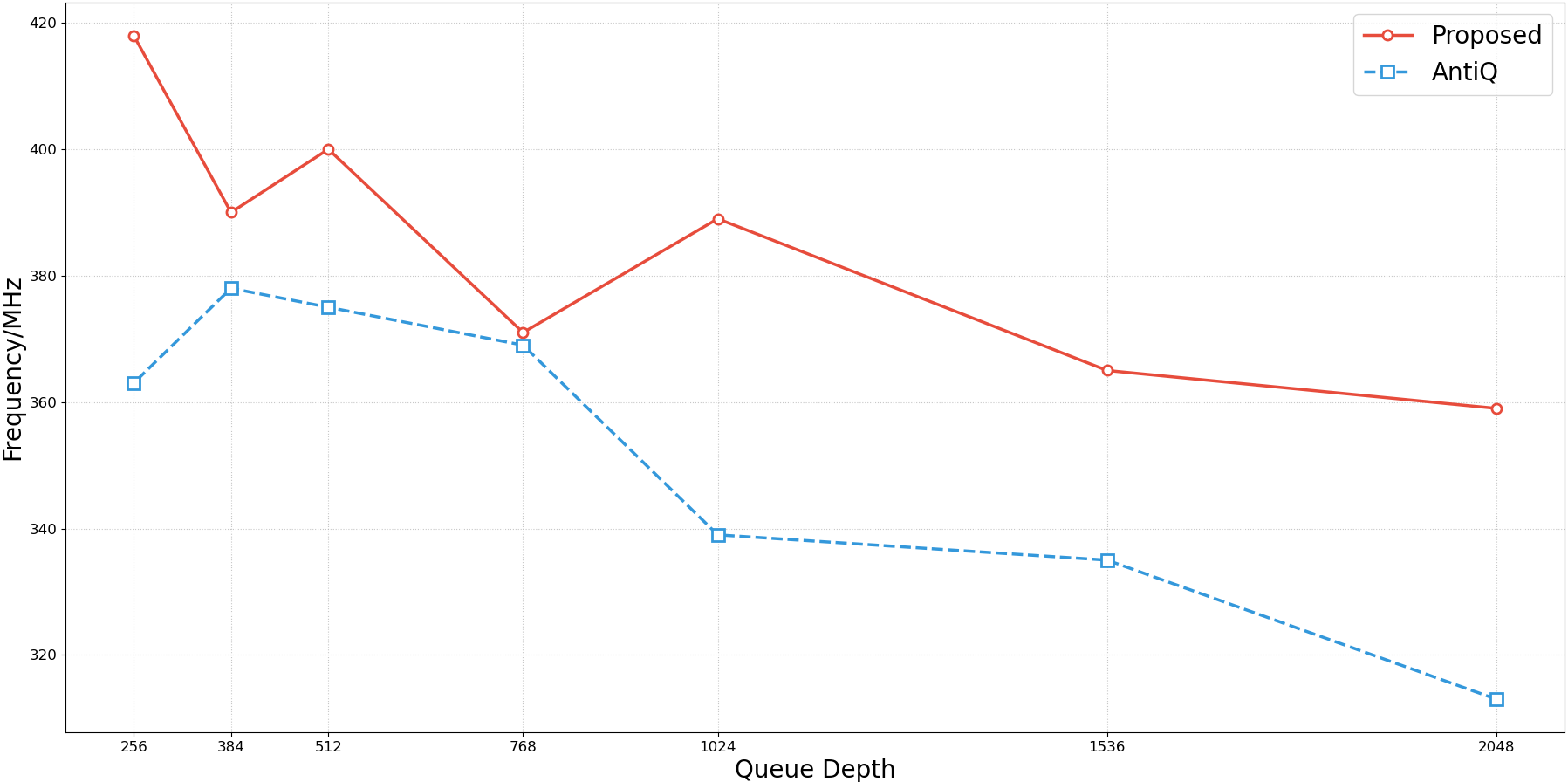}
	}
	\label{fig11:fpga}
	\caption{Comparsion with AnTiQ on FPGA}
\end{figure*}
\subsection{Comparsion}
The open-source timer queue AnTiQ [24] serves as our primary baseline. For a fair comparison, both the ID and DATA widths were configured to 16 bits, supporting 64K unique priorities. To highlight the resource efficiency of our design, we set M=2, which corresponds to the configuration with the largest systolic unit area. We then conducted a comparative analysis against AnTiQ across seven queue depths, ranging from 256 to 2048, evaluating both performance and overhead.\\
The implementation results are presented in Fig. 10. For a given queue depth, our proposed architecture achieves a maximum operating frequency of 418 MHz, outperforming AnTiQ’s 378 MHz. This frequency advantage is consistent across all tested depths. The performance gap is most pronounced at a queue capacity of 256, where our design operates 15\% faster than AnTiQ while
\begin{table}[h]
	\centering
	\label{tab7}
	\caption{Cost and Performance comparsion with AntiQ}
	\begin{tabular}{|c|c|c|c|c|c|c|}
		\hline
		& N & M & Depth & Frequency & LUTs(\%) & FFs(\%)\\
		\hline
		AnTiQ & / & / & 256 & 363 & 3.83\% & 1.28\%\\
		\hline
		Proposed & 128 & 2 & 256 & 418 & 2.54\% & 0.96\%\\
		\hline
		AnTiQ & / & / & 768 & 369 & 11.52\% & 3.86\%\\
		\hline
		Proposed & 384 & 2 & 768 & 365 & 7.63\% & 2.89\%\\
		\hline
	\end{tabular}
\end{table}
consuming 33\% fewer LUTs and 25\% fewer FFs, as detailed in Table VII. Notably, even with M=2—the configuration yielding the largest systolic unit area—our design still demonstrates an average reduction of 31\% in LUT usage and 25\% in FF usage across the seven evaluated queue depths. Table VII further lists the resource disparity at a queue capacity of 768, where the operating frequencies of the two designs are closest. This resource efficiency stems from our architectural choice: AnTiQ employs a Finite State Machine for control within each systolic unit, whereas our design uses a single control logic, implemented with Boolean operations, shared among multiple shift registers within a unit. In terms of operation overhead, AnTiQ requires 3 cycles for a remove and 2 cycles for a enqueue, thus necessitating 5 cycles for an update operation and introducing a potential for operational conflicts. In contrast, our design executes all operations, including update, in a uniform 3-cycle latency. Synthesized under identical conditions (28nm process, same queue depth, and element bit-width), AnTiQ exhibits a stable critical path delay of 3.06 \(ns\). Our architecture maintains a minimal critical path delay of 1.82 \(ns\) at M=2 setting—a reduction of approximately 40\%—and a 13\% smaller area. Furthermore, AnTiQ lacks support for overflow handling. Consequently, from the perspectives of functionality, performance, and resource overhead, our design demonstrates superior advantages.
\section{Conclusion}
To achieve high-precision, large-scale timer management that supports dynamic updates and mitigates timing anomalies from bit-width overflow, we have developed a priority queue architecture. This work introduces two key techniques: an update operation and a group sorting mechanism. The update operation is realized by combining and propagating basic operations to modify the priority of existing elements. The group sorting mechanism ensures correct dequeue timing for expired tasks post-overflow by altering the search sequence. Implemented with a hybrid architecture combining a 1D systolic array and shift registers, our design has been successfully applied to flow table timeout management, enabling hardware dynamic updates and guaranteeing timing accuracy even after overflow.

The integration of update and group sorting capabilities makes this priority queue architecture highly applicable to other dynamic scenarios, such as processor hardware schedulers, dynamic packet scheduling algorithms, and hardware-based anti-starvation mechanisms. In our future work, we aim to reduce resource overhead without sacrificing performance by incorporating RAM macro blocks. Furthermore, we plan to integrate this design into a stateful processing NIC to evaluate its performance within a real-world, high-speed network system.


\begin{IEEEbiography}[{\includegraphics[width=1in,height=1.25in,clip,keepaspectratio]{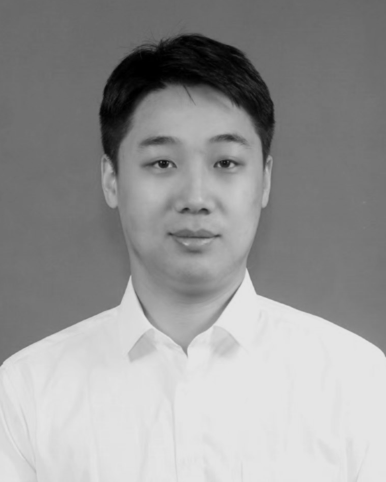}}]{Zekun Wang}
received the B.S. and M.S. degrees in Integrated Circuit Design from Xidian University, Xi’an, China, in 2020 and 2023, respectively. He is currently pursuing the Ph.D. degree in Electronic and Information Engineering at the School of Microelectronics, Xidian University, Xi’an, China. His research focuses on programmable schedulers for network traffic management and Network processor architecture design, with innovations targeting data center network optimization.\end{IEEEbiography}

\begin{IEEEbiography}[{\includegraphics[width=1in,height=1.25in,clip,keepaspectratio]{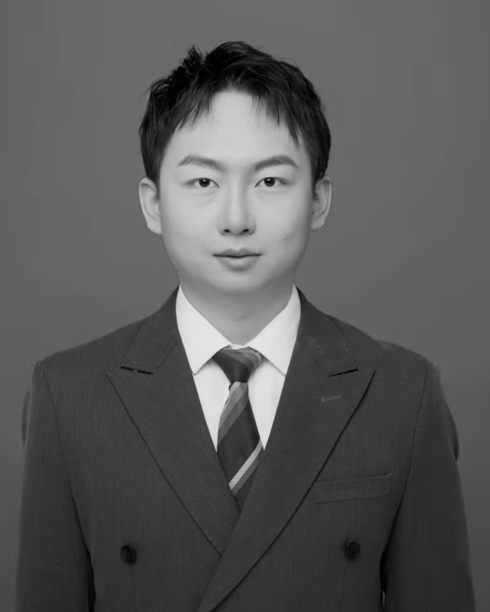}}]{Binghao Yue}
	received the B.S. degree in Communication Engineering from Xidian University, Xi’an, China, in 2023 and is currently pursuing a Master's degree at the School of Communication Engineering, Xidian University. His research focuses on RDMA (Remote Direct Memory Access)-enabled network acceleration  and Smart NIC (System-on-Chip Network Interface Card) architectures, with edge-computing hardware optimization.
\end{IEEEbiography}

\begin{IEEEbiography}[{\includegraphics[width=1in,height=1.25in,clip,keepaspectratio]{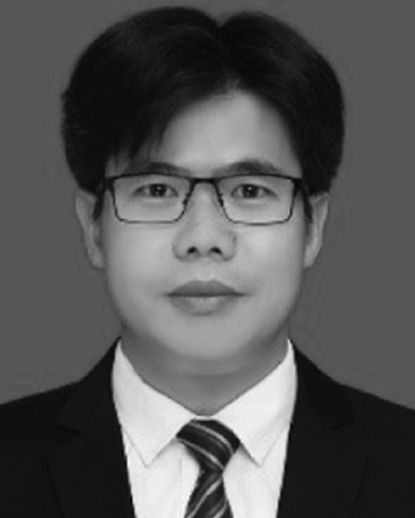}}]{Weitao Pan}
	(Member, IEEE) received the B.S. degree from the School of Technical Physics, Xidian University, Xi’an, China, in 2004, and the Ph.D. degree from the School of Microelectronics, Xidian University in 2010.,He is currently an Associate Professor with the State Key Laboratory of Integrated Service Networks, Xidian University. His current research interests include VLSI design methods and post-silicon verification.
\end{IEEEbiography}

\begin{IEEEbiography}[{\includegraphics[width=1in,height=1.25in,clip,keepaspectratio]{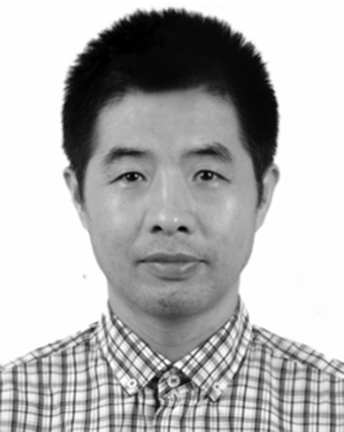}}]{Jiangyi Shi}
received the Ph.D. degree from Xidian University, Xi’an, China, in 2007. He has more than 20 years of experience in IC design, including technical development, computer architecture, and power amplifiers. His research aims at EDA tools and RF design. Recent projects include RF and mixed-signal high-speed circuit design, network processor design, integrated circuit security, and trusted technology.
\end{IEEEbiography}

\begin{IEEEbiography}[{\includegraphics[width=1in,height=1.25in,clip,keepaspectratio]{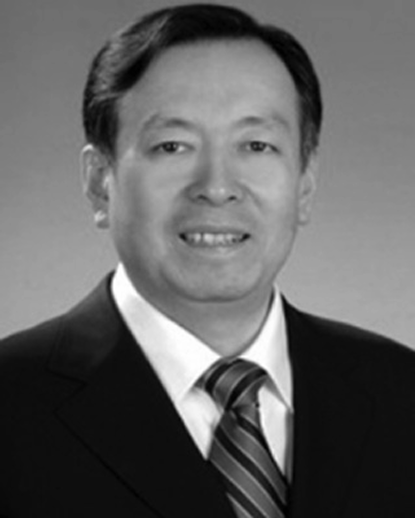}}]{Yue Hao}
was born in Chongqing, China, in 1958. He received the Ph.D. degree from Xi'an Jiao tong University, Xi'an, China, in 1991. He is currently a Professor with State Key Discipline Laboratory of Wide Bandgap Semiconductor Technology, School of Microelectronics, Xidian University, Xi'an. His research focuses on wide forbidden band semiconductor materials and devices.
\end{IEEEbiography}
\end{document}